\documentclass[prl,nofootinbib,twocolumn,superscriptaddress]{revtex4}
\pdfoutput=1
\usepackage{graphicx,rotating}
\usepackage{hyperref}
\usepackage{ifpdf}
\usepackage[english]{babel}
\usepackage{amsmath}
\usepackage{amssymb}

\newcommand{\mrm}[1]{\mbox{\rm #1}}
\newcommand{\be}{\begin{equation}}
\newcommand{\ee}{\end{equation}}
\newcommand{\br}{\begin{eqnarray}}
\newcommand{\bea}{\begin{eqnarray}}
\newcommand{\eea}{\end{eqnarray}}
\newcommand{\er}{\end{eqnarray}}
\newcommand{\ba}{\begin{array}}
\newcommand{\ea}{\end{array}}
\newcommand{\bi}{\begin{itemize}}
\newcommand{\ei}{\end{itemize}}
\newcommand{\bn}{\begin{enumerate}}
\newcommand{\en}{\end{enumerate}}
\newcommand{\bc}{\begin{center}}
\newcommand{\ec}{\end{center}}

\newcommand{\Eq}[1]{Eq.~(\ref{#1})}

\newcommand{\rfn}[1]{(\ref{#1})}

\newcommand{\gev}{\text{GeV}}
\newcommand{\tev}{\text{TeV}}

\newcommand{\gsim}{\lower.7ex\hbox{$\;\stackrel{\textstyle>}{\sim}\;$}}
\newcommand{\lsim}{\lower.7ex\hbox{$\;\stackrel{\textstyle<}{\sim}\;$}}

\newcommand{\hc}[1]{#1^{\dagger}} 

\newcommand{\mdm}{M_{\text{DM}}}
\newcommand{\order}[1]{{\mathcal O}(#1)}

\newcommand{\sdm}{S_{\text{DM}}}
\newcommand{\snl}{S_{\text{NL}}}

\def\mysection#1{{\bf #1.} }

\begin{document}


\title{\bf 
Implications of the CDMS result on Dark Matter and LHC physics 
}

\author{ Mario Kadastik}
\author{ Kristjan Kannike}
\author{ Antonio Racioppi}
\affiliation{National Institute of Chemical Physics and Biophysics, 
Ravala 10, Tallinn 10143, Estonia
} 

\author{ Martti Raidal}
\affiliation{National Institute of Chemical Physics and Biophysics, 
Ravala 10, Tallinn 10143, Estonia
} 
\affiliation{Department of Physics, P.O.Box 64, FIN-00014 University of Helsinki, Finland
}


\vspace*{-2.0 in}

\begin{abstract}
The requirements of electroweak symmetry breaking (EWSB) and correct thermal relic density of Dark Matter (DM) predict 
large DM spin-independent direct detection cross section in scalar DM models based on $SO(10)$ non-supersymmetric
GUTs. Interpreting the CDMS signal events as DM recoil on nuclei, we study implications of this assumption on EWSB, 
Higgs boson mass and direct production of scalar DM at LHC experiments. We show that 
this interpretation indicates relatively light DM, $M_{\text{DM}}\sim \order{100}$ GeV, with large pair production cross
section at LHC in correlation with the spin-independent direct DM detection cross section.
The next-to-lightest dark scalar $S_{\rm NL}$   is predicted to be long-lived, providing distinctive experimental
signatures of displaced vertex of two leptons or jets plus missing transverse energy.
\end{abstract}


\maketitle

\mysection{Introduction}
The existence of cold dark matter (DM) of the Universe
is firmly established by cosmological observations \cite{Komatsu:2008hk}. Because the standard model (SM) of particle interactions 
does not contain a cold DM candidate, its existence is a clear signal of new physics beyond the SM. 
However, the origin, nature and properties of the DM have  so far remained completely unknown. 

The Cryogenic Dark Matter Search (CDMS II)  at Soudan mine has recently observed two 
weakly interacting massive particle (WIMPs) recoil candidate events on nuclei in the signal region \cite{CDMS} and 
additional two events just outside the signal region border \cite{talks}. All the events have recoil energy between 12-15 keV. Although statistically inconclusive, CDMS Collaboration cannot reject the events as 
a signal of  DM scattering on nuclei \cite{talks}. In that case 
 the CDMS II result has important implications on DM theory as well as on DM direct detection at
 colliders. 

Any  theory beyond the SM that attempts to explain the CDMS II result must, in the first place, 
explain what is the WIMP, why it is stable, and to predict correct cosmological DM density 
together with phenomenologically acceptable DM mass scale. After answering those questions one can 
address phenomenological implications of the CDMS II result on other experiments.

It is plausible that the stability of WIMP is due to a discrete $Z_2$ symmetry which 
is an unbroken remnant of some underlying $U(1)$ gauge group \cite{Krauss:1988zc}.
This possibility is particularly attractive because it suggests a Grand Unified Theory (GUT) 
gauge group with a larger rank than that of $SU(5)$ \cite{Georgi:1974sy}.
It was shown in Ref. \cite{Kadastik:2009dj} that if the underlying GUT gauge group is $SO(10)$ \cite{Fritzsch:1974nn},
the argument of  \cite{Krauss:1988zc} predicts that minimal non-supersymmetric DM must be embedded into scalar 
representation ${\bf 16}$  due to the generated matter parity $P_M=(-1)^{3(B-L)}.$ 
In this scenario the DM and its stability mechanism, non-vanishing neutrino masses via the seesaw mechanism \cite{seesaw} and
the baryon asymmetry of the Universe via leptogenesis \cite{Fukugita:1986hr} all spring from 
the same source -- the breaking of  $SO(10)$ gauge symmetry.

The low energy phenomenology of this GUT scenario  is very rich and predictive. 
The DM must consist of dark scalar singlet(s) $S$ \cite{rs} and doublet(s) $H_2$ \cite{id}, thus non-supersymmetric
$SO(10)$ GUT represents an ultraviolet completion  of those scalar DM models \cite{Kadastik:2009dj,Kadastik:2009cu,Frigerio:2009wf}.
The GUT scale boundary conditions, together with the requirements of vacuum stability and perturbativity of scalar self-couplings, 
strongly constrain the allowed parameter space of the theory.
Instead of postulating a negative Higgs $\mu_{1}^{2}$ as in the SM, the electroweak symmetry breaking (EWSB) 
can be dynamically  induced by the Higgs boson interaction with dark scalars either via renormalization group (RG)
evolution of  $\mu_{1}^{2}$ from the GUT scale $M_{\text{G}}$ to $M_Z$~\cite{Kadastik:2009cu} 
as in supersymmetric theories \cite{Ibanez:1982fr}, or via 
DM 1-loop contributions to the Higgs boson effective potential \cite{Hambye:2007vf, Kadastik:2009ca}, 
representing a realistic example of the Coleman-Weinberg idea \cite{Coleman:1973jx}.
Requiring successful radiative EWSB and the observed amount of DM produced in thermal freeze-out, 
the scenario predicts 
the DM spin-independent direct detection cross section to be just at the CDMS II experimental sensitivity.

In this work we argue that CDMS II may have observed the scalar DM recoils on ordinary matter and 
study implications of this fact on the DM direct production processes
at the Large Hadron Collider (LHC) experiments at CERN. Because of the low recoil energy of all the CDMS
events, consistency of CDMS data may indicate relatively light DM detectable at the LHC.
We improve  the Higgs boson effective potential at 1-loop level and require radiative EWSB via renormalization effects. 
We calculate the generated thermal relic DM density and spin-independent direct detection cross section of
DM scattering on nuclei and show that the new CDMS II data prefers light Higgs boson in agreement with 
the precision data analyses \cite{precision}. 
For the obtained parameter space we study the production and decays of dark scalar particles at LHC.
Most importantly for the LHC phenomenology, 
we show that this scenario {\it predicts} long lifetime for the next-to-lightest (NL) neutral dark scalar particle. 
Decays of the NL dark scalar provide an unique 
experimental signature  of displaced vertex in di-lepton and di-quark signal occurs in the  
the decays $S_\text{NL}\to S_{\text {DM}} \ell^+ \ell^-,$ $S_{\text {DM}} q\bar q.$
This signature allows one to discriminate the dark scalar processes over the SM background and
to discover and test light scalar DM at the LHC.

\mysection{The minimal $SO(10)$ scalar DM scenario}
The minimal scalar DM scenario \cite{Kadastik:2009cu} 
contains the SM Higgs in a scalar  representation ${\bf 10}$ and  the DM in a scalar ${\bf 16}$ of $SO(10)$.
Below the  $M_\text{G}$ and above the EWSB scale
the model is described by the $H_{1} \to H_{1}$, $S \to -S,$ $ H_{2} \to -H_{2}$ 
invariant scalar potential
\begin{equation}
\begin{split}
V &= \mu_{1}^{2} \hc{H_{1}} H_{1} + \lambda_{1} (\hc{H_{1}} H_{1})^{2} 
+ \mu_{2}^{2} \hc{H_{2}} H_{2} + \lambda_{2} (\hc{H_{2}} H_{2})^{2} \\
&+ \mu_{S}^{2} \hc{S} S + \frac{\mu_{S}^{\prime 2}}{2} \left[ S^{2} + (\hc{S})^{2} \right] + \lambda_{S} (\hc{S} S)^{2} 
 \label{V}\\
& + \frac{ \lambda'_{S} }{2} \left[ S^{4} + (\hc{S})^{4} \right] 
 + \frac{ \lambda''_{S} }{2} (\hc{S} S) \left[ S^{2} + (\hc{S})^{2} \right] \\
&+ \lambda_{S1}( \hc{S} S) (\hc{H_{1}} H_{1}) + \lambda_{S2} (\hc{S} S) (\hc{H_{2}} H_{2}) \\
&+ \frac{ \lambda'_{S1} }{2} (\hc{H_{1}} H_{1}) \left[ S^{2} + (\hc{S})^{2} \right]
+ \frac{ \lambda'_{S2} }{2} (\hc{H_{2}} H_{2}) \left[ S^{2} + (\hc{S})^{2} \right]
 \\
&+ \lambda_{3} (\hc{H_{1}} H_{1}) (\hc{H_{2}} H_{2}) + \lambda_{4} (\hc{H_{1}} H_{2}) (\hc{H_{2}} H_{1}) \\
&+ \frac{\lambda_{5}}{2} \left[(\hc{H_{1}} H_{2})^{2} + (\hc{H_{2}} H_{1})^{2} \right] \\
&+ \frac{\mu_{S H}}{2} \left[\hc{S} \hc{H_{1}} H_{2} + \mathrm{h.c.} \right]
+ \frac{\mu'_{S H}}{2} \left[S \hc{H_{1}} H_{2} + \mathrm{h.c.} \right], 
\end{split}
\end{equation}
together with the GUT scale boundary conditions
\bea
&\mu_1^2(M_{\text{G}})>0,\; \mu_2^2(M_{\text{G}})=\mu_S^2(M_{\text{G}}) >0, & 
\label{bc1}\\
&\lambda_2(M_{\text{G}})=\lambda_S(M_{\text{G}})=\lambda_{S2}(M_{\text{G}}),\; \lambda_3(M_{\text{G}})=\lambda_{S1}(M_{\text{G}}), \nonumber&
\eea
and
\bea
&{\mu}_S^{\prime 2}, \; {\mu}_{SH}^{ 2} \lsim {\cal O}\left( \frac{M_\text{G}}{M_{\text{P}}}\right)^n \mu^2_{1,2},& 
\label{bc2}\\\
& \lambda_{5} ,\; \lambda'_{S1} ,\; \lambda'_{S2} ,\; \lambda''_{S} \lsim {\cal O}\left( \frac{M_\text{G}}{M_{\text{P}}}\right)^n \lambda_{1,2,3,4}.&
\nonumber
\eea
While the parameters in \Eq{bc1} are allowed by
$SO(10),$ the ones in \Eq{bc2} can be generated only after $SO(10)$ breaking by  operators
suppressed by $n$ powers of the Planck scale $M_{\text{P}}.$

The neutral components of dark bosons yield the scalar mass eigenstates $S_{\text{R}1}$ and $S_{\text{R}2}$, 
and the pseudoscalar ones $S_{\text{I}1}$ and $S_{\text{I}2}$ with  the mass spectrum 
$M_{\text{I}1} \leq M_{\text{R}1} < M_{\text{I}2} \leq M_{\text{R}2},$ (or I$ \leftrightarrow $R) where
$S_{\text{I}1},$ $S_{\text{R}1}$  and $S_{\text{I}2},$  $S_{\text{R}2}$ are almost degenerate in mass due to the smallness of the parameters in
\Eq{bc2}.  For clarity we denote the DM particle by $\sdm$ and 
the NL scalar by $\snl$. 

We stress that the mass degeneracy of $\sdm$ and $\snl$ is a generic {\it prediction} of the scenario and follows from 
the underlying $SO(10)$ gauge symmetry via \Eq{bc2}. This degeneracy has several phenomenological implications
which allow to discriminate this scenario from other DM models. First, it implies long lifetime for 
 $\snl$ which provides clear experimental signature of displaced vertex in the decays $S_{\text{NL}}\to S_{\text {DM}} \ell^+ \ell^-$
 at the LHC. We study this experimental signature in this work. Second, it offers a possibility to 
 reconcile the DAMA/NaI and DAMA/LIBRA annual modulation signal \cite{Bernabei:2009du} 
 with the results of XENON10 \cite{Angle:2007uj} and CDMS II \cite{CDMS}
 experiments via the idea of inelastic DM scatterings \cite{TuckerSmith:2001hy}. 
 The inelastic DM requires degenerate mass states as predicted by \rfn{bc2}.
 While spin-dependent inelastic scatterings may explain all the data \cite{Kopp:2009qt}, 
 the spin-independent DM scatterings as a solution to DAMA
 signal is (almost) excluded by XENON10 and CDMS II results  \cite{Fairbairn:2008gz}, and 
 we do not pursue this possibility here.
 
 At $M_{\text{G}}$ the SM gauge symmetry is not spontaneously broken, $\mu_1^2(M_{\text{G}})>0.$
 To obtain EWSB at low energies we require that $\mu_1^2(M_Z)$ becomes negative 
 by RG evolution \cite{Kadastik:2009cu} which is the only possibility  in the case of light DM. 
 We RG improve  \cite{Casas:1998cf} our previous calculation with 1-loop corrections to the effective potential
 as described in \cite{Kadastik:2009ca}.


\mysection{DM direct detection and Higgs boson mass}
In our scenario both the DM annihilation at early Universe and the DM scattering on nuclei  
 are dominated by  tree level SM Higgs boson exchange. 
The relevant  DM-Higgs effective coupling  is
\begin{equation}
\lambda_{\text{eff}} \, v = \frac{1}{2} (\sqrt{2} s\, c\, \mu'_{SH} - 2 s^{2} (\lambda_{3}+\lambda_{4}) v -2 c^{2} \lambda_{S1} v),
\label{lameff}
\end{equation}
where $s,\,c$ are the sine, cosine of the singlet-doublet mixing angle. 
We systematically scan over the full parameter space of the model by iterating between $M_\text{G}$ and $M_Z$ 
using RGEs of Ref.  \cite{Kadastik:2009cu}.  We require successful dynamical EWSB and calculate the thermal freeze-out DM abundance and 
spin-independent direct detection cross section per nucleon using MicrOMEGAS package~\cite{micromegas}. 
The latter is approximately given by
\be
 \sigma_{\text{SI}} \approx \frac{1}{\pi} f_N^2 \left( \frac{\lambda_{\text{eff}}v}{v M_{\text{DM}}}  \right)^2 \left( \frac{M_N}{M_h} \right)^4 ,
 \label{sigma}
\ee
where $f_N\approx 0.47$ is the nucleonic form factor that
includes all contributions from the valence and sea quarks ($s$-dominated) and gluons.

\begin{figure}[t]
\includegraphics[width=0.37\textwidth]{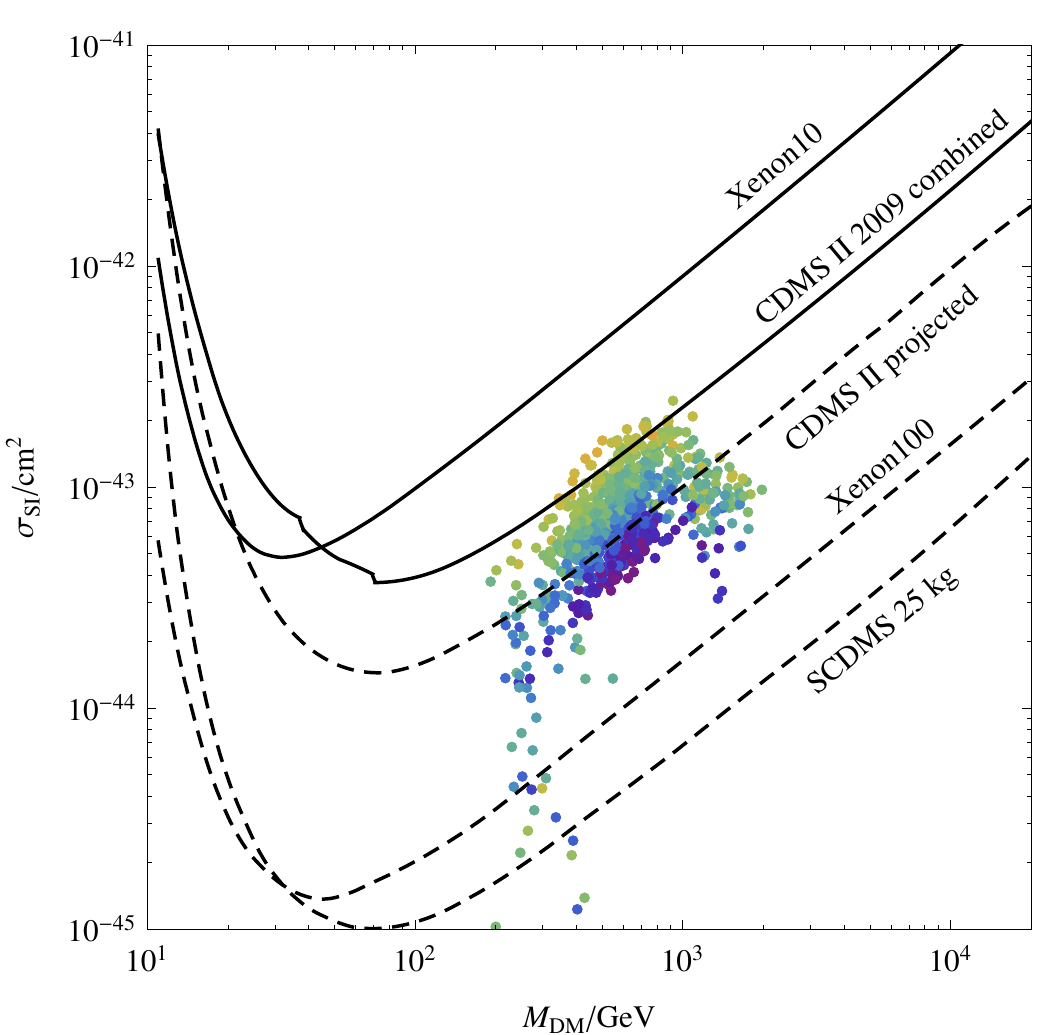}
\caption{DM direct detection cross-section/nucl.~{\it vs}.~$M_{\mathrm{DM}}$. 
Color shows SM Higgs boson masses from 130~GeV (yellow) to 180~GeV (violet). 
Solid lines represent current bounds, dashed lines are expected future sensitivities.
}
\label{fig:dd}
\end{figure}

We present in Fig.~\ref{fig:dd} our prediction for the spin-independent DM cross section as a function of DM mass
for different Higgs boson masses described by the colour code $130~\mrm{GeV}<M_h<180~\mrm{GeV}$ from yellow
to violet. The present experimental bounds on $ \sigma_{\text{SI}} $ (solid lines) \cite{CDMS,Angle:2007uj} together with 
the expected sensitivities (dashed lines) are also shown.
For every point we require the WMAP 3$\sigma$ result $0.094 < \Omega_{\text{DM}} < 0.129.$  
For the high mass points, $M_{\text{DM}}>700$ GeV,  EWSB is obtained radiatively via effective potential due to large 
 ``soft" portal coupling $\mu'_{SH}$  \cite{Kadastik:2009ca}.
This parameter region cannot be directly tested at the LHC.
In the low mass region, $180~\mrm{GeV} < M_{\text{DM}}<700$ GeV, EWSB occurs due to renormalization of $\mu^2_1$
from $M_{\text{G}}$ to $M_Z.$ The lower bound on the DM mass comes from the top quark loop
contributions to the SM Higgs boson effective potential. The low mass points  in Fig.~\ref{fig:dd}  
with small $\sigma_{\text{SI}}$ are due to cancellations between the parameters in \Eq{lameff}.
If the CDMS II events \cite{CDMS} indeed are DM recoil,   those points are excluded.
 Fig.~\ref{fig:dd} clearly prefers light Higgs boson to increase the 
 spin-independent cross section \rfn{sigma} to explain the CDMS~II events.


\mysection{ LHC phenomenology}
The main parton level production processes for the dark sector scalars at the LHC are 
 $q \bar{q} \to H^{+} H^{-}$,  $g g \to S_{\text{DM},\text{NL}} S_{\text{DM},\text{NL}}$, 
$q \bar{q}' \to S_{\text{DM},\text{NL}} H^{\pm}$  and $q \bar{q} \to \snl \sdm,$
followed by the decays  $S_{\text{NL}}\to S_{\text{DM}} \ell^+ \ell^-,$ $ S_{\text{NL}}\to S_{\text{DM}}q\bar q$
and $H^+\to S_{\text{DM}} q \bar q,$ $H^+\to S_{\text{DM}} \ell \bar\nu.$
Although the pair production $H^+H^-$ is the least model dependent process due to the virtual $\gamma, $ $Z$ exchange, the
cleanest experimental signature is provided by the $\snl$ decays.  
Since the DM and the NL scalar are almost degenerate in mass, the mean lifetime $\tau$ of NL scalar can be long
and $\snl$ can travel macroscopic distances before decaying. Therefore the decays $S_{\text{NL}}\to S_{\text {DM}} \ell^+ \ell^-,$
$S_{\text {DM}} q\bar q$ can be tagged by the displaced vertex of lepton or jet pairs and missing transverse energy $E_{\text{T}}$, providing
SM background free signal of scalar DM at the LHC.

\begin{figure}[t]
\includegraphics[width=0.4\textwidth]{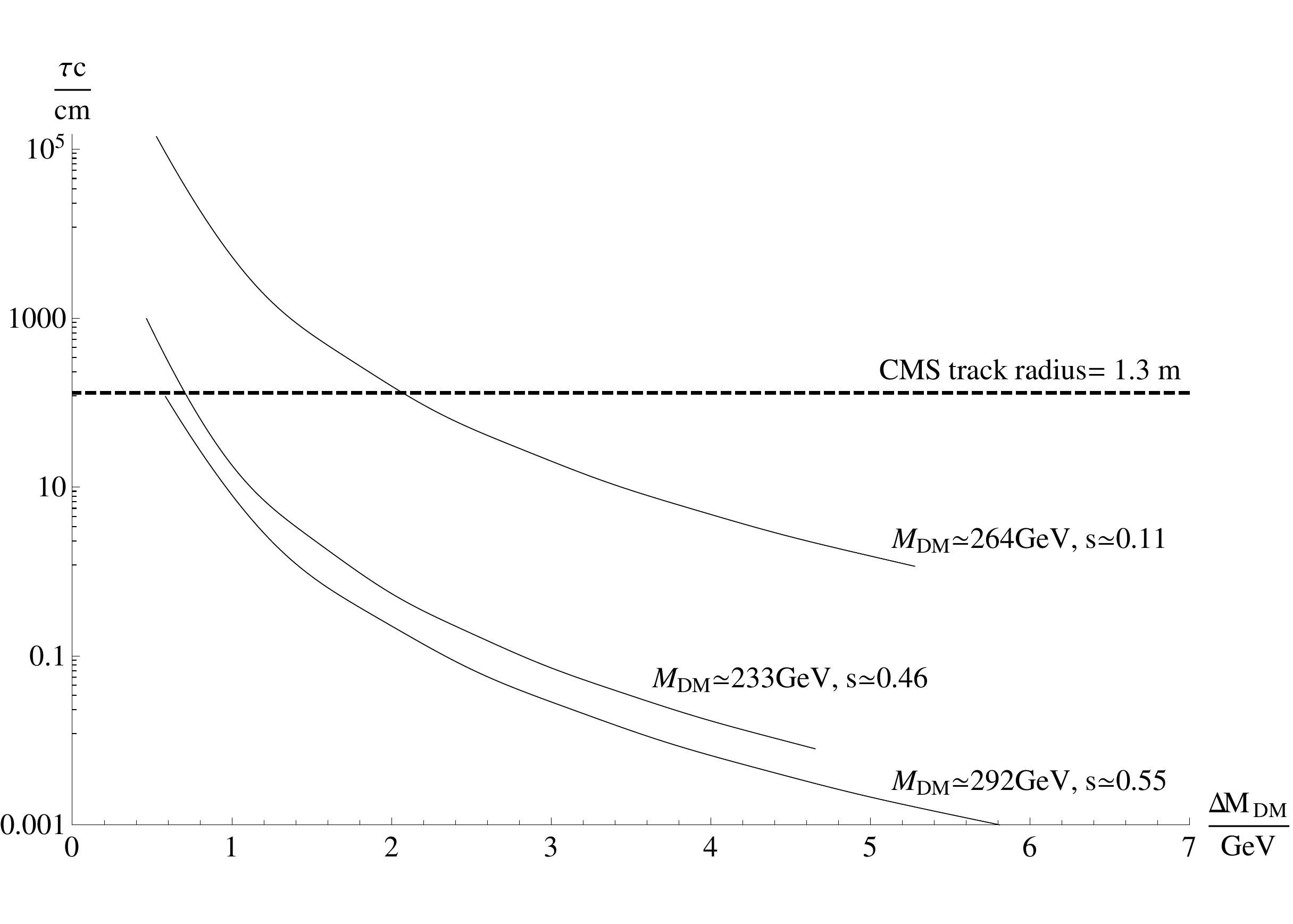}
\caption{ Distance of DM displaced vertex from the interaction region at LHC experiments as a function of $\Delta \mdm.$  
Examples for three data points with different values of DM mass $\mdm$ and sine of the mixing angle $s$ are show. 
The dashed line is the CMS tracker radius \cite{:2008zzk}.
}
\label{fig:dp:vertex}
\end{figure}

In Fig. \ref{fig:dp:vertex} we plot the distance $c \tau$ of the displaced vertices  from the  interaction region
as a function of the mass splitting 
$\Delta \mdm=M_{\text{NL}}-\mdm$ for three examples of DM mass and  mixing. 
Depending on the mass gap and the mixing angle, the displacement can range from 
micrometers to several meters. In the case of such a \emph{far} displaced vertex, 
the experimental signatures of the $\snl\snl$ final state are $\ell\ell\ell\ell,$ $jjjj,$ $\ell\ell j j$ with displaced 
$\ell\ell$ or $jj$ vertices and the missing $E_{\text{T}}.$ The decays with $\nu\nu$ final states are 
completely invisible, but those make up roughly 20\% of the total $\snl$ branching fraction depending on $\Delta \mdm .$ 
Therefore we predict high efficiency in the detection of NL scalar decays at the LHC.   

We have computed the $\snl\snl,$ $\sdm\sdm,$ $\sdm\snl$ and  $H^{+} H^{-}$ production cross sections $\sigma_{\text{LHC}}$
in $pp$ collisions at LHC by convoluting over the parton distribution function of Ref. \cite{Alekhin:2006zm}. 
We note that the production processes  are similar to those of the Inert Doublet model studied in \cite{Cao:2007rm}, 
but the dependence on
model parameters is very different in our case. 

\begin{figure}[t]
\includegraphics[width=0.4\textwidth]{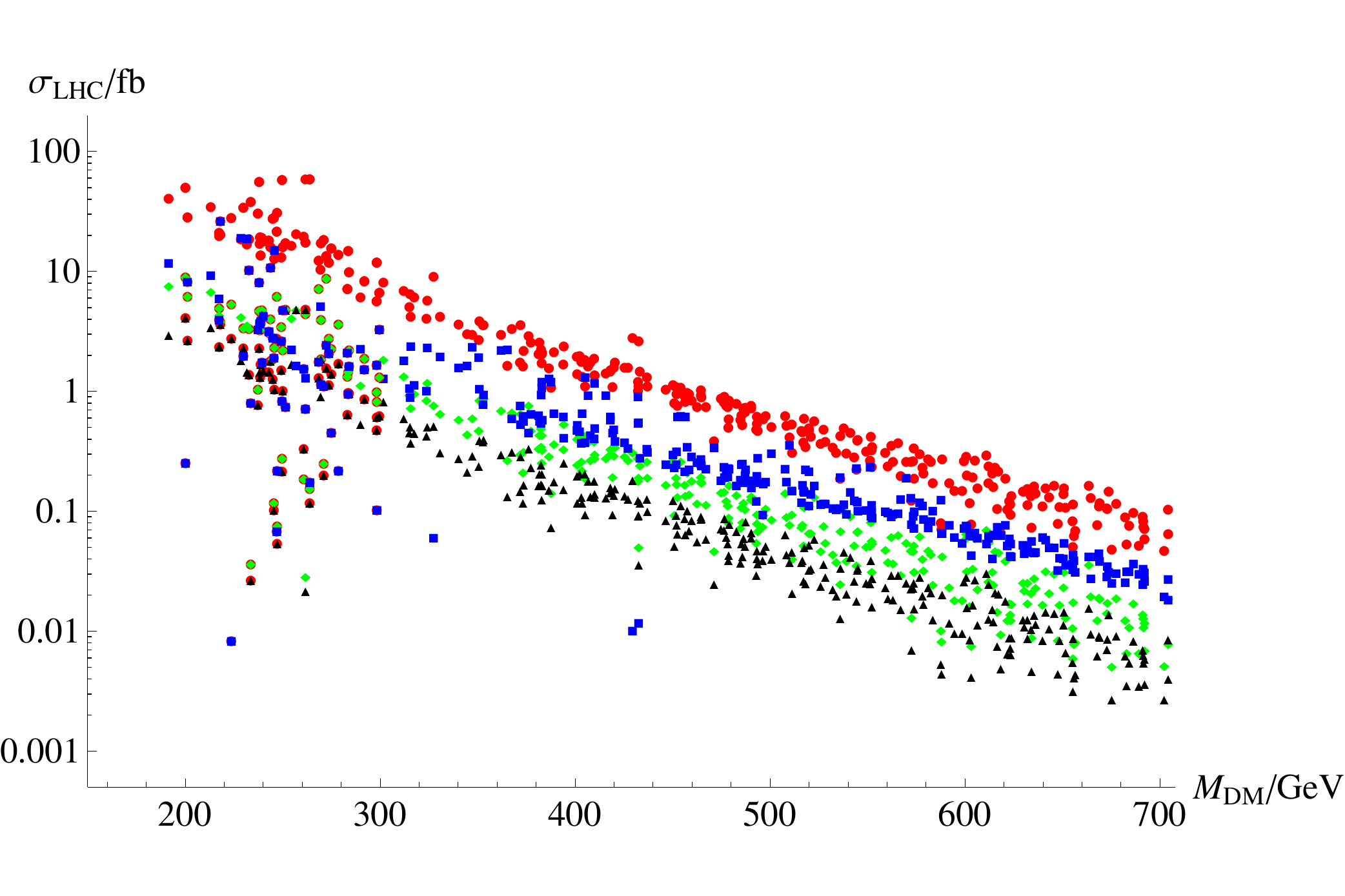}
\caption{Direct production cross-section 
of $pp \to H^{+} H^{-}$ (red), $pp \to S_{\text{DM},\text{NL}} S_{\text{DM},\text{NL}}$ (blue), 
$pp \to S_{\text{DM},\text{NL}} H^{\pm}$ (green) and $pp \to \snl \sdm$ (black) at the LHC for $\sqrt{s} = 14~\tev$. 
}
\label{fig:cs:lhc}
\end{figure}

In Fig. \ref{fig:cs:lhc} we show the cross sections for these processes for the collision energy $\sqrt{s} = 14~\tev$ as a function of DM mass
for {\it the same points}  as in Fig.~\ref{fig:dd}. The colour code is explained in the caption. Because $\sdm$ and $\snl$ are almost degenerate,
their production cross sections are almost equal.
Because the DM particle is neutral, the $H^{+}$ cross section exceeds the $\sdm$ and $\snl$ ones. 
The cross sections fall rapidly with outgoing particle masses. Taking into account the visible branching fractions as described
above, DM can be discovered at the LHC for $\mdm < {\cal O}(300)~\gev$.

Comparison of Fig.~\ref{fig:dd} and Fig. \ref{fig:cs:lhc} reveals correlation between the low $\mdm$ points with accidentally small
$\sigma_{\text{SI}}$ in Fig.~\ref{fig:dd}  and the ones with small  $\sigma_{\text{LHC}}$ in Fig. \ref{fig:cs:lhc}.
Interpreting the CDMS II events as the DM recoil excludes the small cross sections regions.
Therefore the CDMS II result implies large DM production cross section at LHC provided DM  mass is below $\mdm <  {\cal O}(300)~\gev$.
The recoil energy 12-15 keV of the
CDMS events puts a lower bound on the DM mass, $M_{\text{DM}} >{\cal O}(10)$ GeV.
Clearly the consistency of  recoil energy of all the CDMS II events does not exclude the possibility of heavy DM 
but it favours light DM, $\mdm \sim{\cal O}(100)~\gev$.
Therefore, if the observed CDMS events really are due to DM recoils, our results show 
that scalar DM can be directly discovered at the LHC.

\mysection{Conclusions}
We have considered implications of the recent CDMS II data on the minimal scenario of $SO(10)$ scalar DM. 
The parameter space of the model is strongly constrained by the requirements of vacuum stability, perturbativity, correct DM density and 
dynamical  EWSB via DM interactions, {\it cf.}  Fig.~\ref{fig:dd}.
If we assume that the observed signal events in the CDMS II experiment were DM recoil on nuclei, 
the recoil energy and the position of the highest sensitivity region for the CDMS II suggest 
a light DM with large spin-independent cross section, which implies 
 large  DM production cross section at the LHC.
The distinctive collider signature of this scenario is  a highly displaced vertex 
of two leptons or jets  and missing transverse energy. 
The Higgs boson must be light to explain the CDMS II observations.

\vskip 0.3 cm

\noindent {\bf Acknowledgment.}
We thank  S. Davidson and  K. Huitu  for discussions.
This work was supported by the ESF 8090, JD164,  SF0690030s09 and  EU FP7-INFRA-2007-1.2.3 contract No 223807.


\end{document}